\documentclass[amssymb,superscriptaddress,twocolumn,prd]{revtex4}
\usepackage{graphicx}
\def \be{\begin{equation}}
\def \ee{\end{equation}}
\def \bea{\begin{eqnarray}}
\def \eea{\end{eqnarray}}

\newcommand{\beq}[1]{\begin{eqnarray}\label{#1}}
\newcommand{\eeq}{\end{eqnarray}}

\begin{document}
 \pagestyle{plain}

 \title{Spherical Collapse Model And Dark Energy(I)}

\author{Ding-fang Zeng and Yi-hong Gao}
\email{dfzeng@itp.ac.cn} \email{gaoyh@itp.ac.cn}
\affiliation{Institute of Theoretical Physics, Chinese Academy of
Science.}
 \begin{abstract}
 In existing literatures about the top-hat spherical collapse model of
 galaxy clusters formation in cosmology containing dark energies,
 dark energies are usually
 assumed not to cluster on this scale. But all these literatures
 ignored the current describing the flowing of dark energies
 outside the clusters which should exist under this
 assumption, so the conclusions of these literatures are worth
 further explorations. In this paper we study this model in QCDM or Phantom-CDM
 cosmologies(flat) by assuming that dark energies will cluster
 synchronously with ordinary matters on the scale of galaxy clusters
 so the dark energy current flowing outside the clusters
 does not exist at all and find that in this case, the
 key parameters of the model exhibit rather non-trivial and
 remarkable dependence on the equation of state coefficients of
 dark energies. We then apply the results in
 Press-Scheter theory and calculate the number density of galaxy
 clusters and its evolutions. We find that this two quantities are both
 affected exponentially by the equation of
 state coefficients of dark energies. We leave the study of this
 model with the assumption that dark energies do not cluster on
 the scale of galaxy clusters at all as the topic of another paper
 where similar conclusions will be obtained also.
 \end{abstract}


 \maketitle

 \section{Introduction}

 When Press-Schechter theory is used to study the number density of
 galaxy clusters and its evolutions, there are two key parameters
 \cite{PeeblesBigBook}, \cite{Dodelson} needing to be extracted
 from the spherical collapse models. \cite{GunnGott, LaceyCole}
 are two early works studying this model. They considered
 it in a background universe without dark energies.
 \cite{Barrow, ECF, WangSteinhardt1, LokasHoffman, Peacock, NWeinberg}
 are some recent works containing discussions about this
 model in $\Lambda$CDM or QCDM cosmologies. In these existing
 works, when dark energy is included, it is assumed that it does not
 cluster on the scale of galaxy clusters.

 However, if dark energy does not cluster on this scale, then
 when we write down the Einstein equations to describe the
 evolution of an over-dense region which will develop into
 the future galaxy clusters, we should
 include a dark energy current in the energy momentum tensor
 to describe the dark energy component flowing outside
 the over-dense region. But, in all these existing literatures,
 such a current was ignored.
 To see the effects of such a current, there are two methods. The
 first is, adding it in the energy momentum tensor and solving the
 resulting Einstein equations. The second is, assuming that such a
 current does not exist at all and solving the appropriate
 Einstein equations to see the effects of dark energy's clustering
 on the scale of galaxy clusters.

 The purpose of this paper is to study this problem by the second
 method. We leave the study of the first way as the topic of
 another paper \cite{SphereII}, in that paper we will get similar results
 as in this paper. After a simple introduction to Press-Scheter theory in
 section \ref{PS_section}, we study the spherical collapse model
 in section \ref{SCMinQCDM_section} in details. Our study
 shows that when dark energy is assumed to cluster synchronously
 with the ordinary matters so the dark energy current does
 not exist on the scale of galaxy clusters, the key
 parameters of the model exhibit rather non-trivial and
 remarkable dependence on the equation of state coefficients of
 dark energies. We then apply
 the results obtained in this section in
 Press-Scheter theory and calculate the number
 density of galaxy clusters and its evolutions in
 section \ref{PS_App_Section}. We find
 that, under this assumptions, the number density of galaxy
 clusters as well as its evolutions are both affected
 exponentially by the equation of state coefficients of
 dark energies. We then study the possibility of determining
 the equation of coefficients of dark energies by measuring
 the number density of galaxy clusters. We end the paper in section \ref{ConSection}
 which contains the main conclusions of this paper
 and some discussions.

 Our study is performed for time-independent equation of
 state coefficients of dark energies $w$ which ranges
 in $-\infty\sim0$. But in
 some cases we will use terminology Quintessence or
 Phantoms regardless how much $w$ is.

 \section{Press-Schechter Theory}\label{PS_section}

 Press-Schechter Theory \cite{PressSchechter} predicts that the
 fraction of volume which has collapsed at a certain red-shift $z$ is
 \beq{}
 f_{coll}(M(R),z)=\frac{2}{\sqrt{2\pi}\sigma(R,z)}
 \int_{\delta_c}^{\infty}d\delta e^{-\delta^2/2\sigma^2(R,z)}.
 \nonumber\\
 \label{Press-Schechter}
 \eeq
 Here, $R$ is the radius over which the density field has been
 smoothed, $\sigma(R,z)$ is
 the rms of the smoothed density
 field \cite{Dodelson}. $\delta_c$ is the threshold of density contrast at time
 $z$ (count stopping time)
 beyond which objects collapse.

 By the notations of
 \cite{Dodelson}, chapter 7,
  \beq{}
 &&\hspace{-3mm}\sigma^2(R,z)=\nonumber\\
 &&\hspace{-3mm}\sigma_8^2\cdot\frac{\left[\int k^{n_s+2}T^2(k)W^2(k\cdot R)dk\right]D_1^2([1+z]^{-1})}
 {\left[\int k^{n_s+2}T^2(k)W^2(k\cdot 8h^{-1}\textrm{Mpc})dk\right]D_1^2(1)}
 \label{sigmaRz}
 \eeq
 where $n_s$ is the spectral index of perturbation powers,
 $n_s=1$ corresponds to scale invariant power; $T(k)$ is the
 transfer function, in this paper we will use the BBKS fitting
 formulaes for it \cite{BBKS, Sugiyama, WangSteinhardt1}; $W(k\cdot R)$ is the
 smoothing window function, we will use the top hat window in this
 paper; $D_1([1+z]^{-1})$ is the
 growth function of linear perturbation theory,
 \beq{}
 D_1(a)=\frac{5\Omega_{m0}H^2_0}{2}H(a)\int_0^ada^{\prime}[a^{\prime}H(a^{\prime})]^{-3}.
 \eeq

 Operationally \cite{ECF},
 $\delta_c$ is defined as the extrapolation of primordial
 perturbations to the collapse epoch using the growth law of
 linear perturbation theory, i. e.,
 \beq{}
 \delta_c=\left[(\frac{\rho_{mc}(a)}{\rho_{mb}(a)}-1)\frac{1}{D_1(a)}\right]_{a\rightarrow0}D_1(a_c),
 \label{deltacDefinition}
 \eeq
 where $\rho_{mc}$ and $\rho_{mb}$ are the matter densities of
 cluster and background respectively.
 It can be shown that $D_1(a)_{a\rightarrow0}\rightarrow a$. Using the method of
 \cite{LaceyCole} we can show that
 \beq{}
 \left[\frac{a_p}{a}\right]_{a\rightarrow0}=(1-\alpha\cdot a),
 \label{roaagoesto0}
 \eeq
 where $a_p$ is the scale factor of the cluster, we use the
 normalization for $a_p$ so that when $a\rightarrow0$, $a_p\approx
 a$, see section \ref{MetricAnsaltz}. Substituting
 eq(\ref{roaagoesto0}) into eq(\ref{deltacDefinition}) we get
 \beq{}
 \delta_c=3\alpha\cdot D_1(a_c).
 \label{deltacOperationDefinition}
 \eeq
 By studying spherical collapse
 model we will give an analytical formulae for the calculation of $\alpha$.

 We comment here that, because $D_1(a)_{a\rightarrow0}\rightarrow a$,
 as long as the definition eq(\ref{deltacDefinition}) is to be of any
 sense, we must have $[\frac{\rho_{mc}}{\rho_{mb}}-1]_{a\rightarrow0}\sim a$. So
 eq(\ref{roaagoesto0}) holds regardless whatever the background
 cosmology model is. Physically, this is because, any realistic
 cosmological model behaves like a flat matter dominated cosmology
 at early times, so they have the same limit behavior as $a\rightarrow0$. Mathematically,
 eq(\ref{roaagoesto0}) is only derived out explicitly in the
 OCDM cosmologies \cite{LaceyCole} and $\Lambda$CDM cosmologies
 \cite{LokasHoffman} and \cite{Peacock}.

 It should be noted besides the partition of the
 cosmological components and the observational epoch, the value
 of $\delta_c$ is also quite dependent on the choice of
 smoothing window used to obtain the dispersion
 $\sigma(R,z)$ \cite{LaceyCole}. We will not consider
 this effect in this paper.

 \section{Spherical Collapse Model in QCDM Cosmologies}\label{SCMinQCDM_section}

 \subsection{Einstein Equations and Energy Conservation}\label{MetricAnsaltz}

 In \cite{WangSteinhardt1} and \cite{NWeinberg}, it is stated that
 the energy hold by the dark energy-Quintessence inside the the over-dense region
 varies independently of the region's radius, and the
 curvature parameter of the over-dense region is time dependent,
 so we have no Friedman-like equation, with a constant curvature
 parameter to describe the evolution of the over-dense region.
 However, if this is the case, then when we write down the Einstein
 equation $G_{\mu\nu}=-8\pi GT_{\mu\nu}$, we should include a
 dark energy current -Q current- in the energy momentum tensor to describe
 the flowing of dark energy outside the over-dense region. But
 \cite{WangSteinhardt1} and \cite{NWeinberg} did not do so.

 To see the effects of such a current, we have two choices. The first is
 directly including such a current in the energy momentum tensor
 and solving the resulting Einstein equations, we will do this
 way in \cite{SphereII}. The second is
 assuming that such a Q-current does not exist and solving the
 resulting Einstein equations to see the effects of Quintessence's
 clustering on the properties of galaxy clusters.
 In this method, Quintessence
 will cluster or expand synchronously with matters
 in the over-dense region, just as we usually
 do in the case of universes. This is an indirect
 but easier method. We will take this way in this paper.

 Considering an uniform over-dense region in the back ground of a flat cosmology
 containing general Quintenssence component, the background
 satisfies
 \beq{}
 (\frac{\dot{a}}{a})^2=\frac{8\pi G}{3}(\rho_{mb}+\rho_{Qb}).
 \label{FriedmanBackground}
 \eeq
 For the over-dense region, starting from the metric ansaltz
 \beq{}
 ds^2=-dt^2+U(t,r)dr^2+V(t,r)(d\theta^2+\textrm{sin}^2\theta
 d\phi^2),
 \label{metric1}
 \eeq
 using Einstein equations, we can prove that
 \beq{}
 U(t,r)=\frac{a_p^2(t)}{1-kr^2}, V(r)=a_p^2(t)r^2,\label{UVfunction}
 \eeq
 where $k$ is a constant independent of time while $a_p$ satisfy:
 \beq{}
 a_p\ddot{a}_p=-\frac{4\pi
 G}{3}[\rho_{mc}+\rho_{Qc}+3p_{Qc}]a_p^2,
 \label{time_time_Einstein}
 \eeq
 \beq{}
 a_p\ddot{a}_p+2\dot{a}_p^2+2k=4\pi
 G[\rho_{mc}+\rho_{Qc}-p_{Qc}]a_p^2.\label{space_space_Einstein}
 \eeq
 The difference of eqs(\ref{time_time_Einstein})
 and (\ref{space_space_Einstein}) yields
 \beq{}
 [\frac{\dot{a}_p}{a_p}]^2+\frac{k}{a_p^2}=\frac{8\pi
 G}{3}\rho(t).
 \label{Friedmann_eq}
 \eeq
 In eq(\ref{FriedmanBackground}), $a$ denotes the scale factor of
 the background universe, $\rho_{mb}$ and $\rho_{Qb}$ denote
 the density of matter and Quintessence in it respectively. While in
 eq(\ref{UVfunction})-(\ref{Friedmann_eq}), $a_p$ denotes
 the scale factor of the over-dense region, the future galaxy
 clusters. $\rho_{mc}$, $\rho_{Qc}$ denote the density of matter
 and quintessence inside the over-dense region respectively, while
 $p_{Qc}$ denotes the pressure produced by the Quintessence.
 By the statement of \cite{WangSteinhardt1} and \cite{NWeinberg},
 the curvature
 parameter $k$ of the over-dense region is time-dependent. This is
 equivalent to say, the metric function $U(t,r)$ cannot be
 factorized as eq(\ref{UVfunction}). If this is the case, then
 not only Freidmann equation
 (\ref{Friedmann_eq}), but also Raychaudhuri-equation
 (\ref{time_time_Einstein}), or eq(A2) of \cite{WangSteinhardt1}
 will follow the usual way.

 In practice, we need taking a normalization for the definition of
 scale factor $a_p$ of the over-dense region. We choose the
 convention so that $a_p\approx a$ when $a\rightarrow 0$.
 Under this convention, the value of $k$ in
 eq(\ref{UVfunction}) is not 1 definitively, its value
 depends on the total energy density inside the over-dense region.
 But, it is a constant independent of time.

 Using the metric ansaltz eq(\ref{metric1}) and
 (\ref{UVfunction}), the energy conservation
 equation $T^\mu_{\nu;\mu}=0$ can be integrated to give
 \beq{}
 \rho_ma^3_p+\rho_Qa_p^{3(1+w)}=\textrm{a\ constant},
 \label{QCDMconservation}
 \eeq
 where $w$ is the equation of state coefficients of the
 Quintessence($-1<w<0$) or phantoms($w<-1$). In
 this paper, we only consider Quintessence/phantoms with
 time-independent $w$ models.
 During the formation of galaxy clusters, the mass of the system
 is conserved at least approximately. This directly leads to the
 conclusion that, the density of Quintessence inside the
 over-dense region does not vary independently of the radius of
 the region. If the equation of state coefficients satisfy
 $-1<w<0$, then as $a_p$ decreases,
 $\rho_{Qc}$ increases, Quintessence clusters. If $w<-1$,
 then as $a_p$ decreases, $\rho_{Qc}$ also decreases, i.e.,
 phantom is repulsed outside the over-dense region, it
 anti-clusters. Contrary to our reasonings here, in \cite{NWeinberg},
 the energy held by Quintessence in the over-dense region does not
 conserve, see eq(14) of it.

 In \cite{CDS1,CDS2,CDS3}, it is pointed out that on cosmological scales,
 a smoothly distributed,
 time-varying component violates equivalence
 principle, so is un-physical. Here, on the scale of galaxy clusters,
 we see that to preserve Einstein equations, we should either
 include a Q-current in the energy momentum tensor, or assume that
 Quintessence (or phantom, even time independent vaccum energy)
 clusters (anti-clusters) just like
 matters. As is well known, when the total
 universe is looked as a cluster of size Hubble scale,
 such a current which describes Q-component flowing outside
 the Hubble horizon is usually assumed not exist.
 Galaxy clusters are a class of objects whose formation and
 evolution is still in the linear perturbation theory applicable
 area and in the energy composition of these objects, dark energy
 occupies a rather big part, just like our biggest cluster - our
 total universe. So the
 assumption that Quintessence will cluster on the scale of galaxy
 clusters like ordinary matters is at least partly the case. We
 will also see in \cite{SphereII} that, if dark energy is assumed not to cluster on
 the this scale, then all the effects displayed in this paper will
 be enforced instead of weakened.

 \subsection{The Key Parameters of the Spherical Collapse Model of Galaxy Clusters Formation}

 In the ideal model, if there is an over-dense region in a flat
 background universe, then at very early times, this region will
 expand as the background universe expands; but because this
 region's over-dense, its expanding rate will decrease and stop
 doing so at some middle times; then it starts to shrink because
 of self-gravitating, the final fate of this over-dense region
 is a singular point. But in practice, when this region shrinks to
 some degree, the pressures originate from the random moving of
 particles inside the over-dense region will balance the
 self-gravitation and the system enters the virialization
 period. In theoretical studies, it is usually assumed that the
 virialization point is coincident with the collapse point of
 the ideal model on the time axis.

 According to Press-Shceter theory eq(\ref{Press-Schechter}),
 if an over-dense region is to
 be virializated at some time $a_c$, its density-contrast should be
 no less than $\delta_c(w,\Omega_{m0},a_c)$. While to relate the
 mass of a galaxy cluster with its characteristic X-ray temperature,
 the ratio of cluster/background matter densities at
 the virialization point is a very important parameter,
 \beq{}
 \Delta_c(w,\Omega_{m0},a_c)=\frac{\rho_{mc,c}}{\rho_{mb,c}}.
 \label{DeltacDefinition1}
 \eeq

 To see the effects of Quintessence/Phantom's
 clustering/anti-clustering on the
 formation and properties of the galaxy clusters, we would like to
 calculate this two important parameters in this section. To
 calculate $\delta_c(w,\Omega_{m0},a_c)$, starting from
 eq(\ref{FriedmanBackground}) and (\ref{Friedmann_eq}), let
 \beq{}
 &&\zeta=\frac{\rho_{mc,ta}}{\rho_{mb,ta}}
 \label{zetaDefinition}\\
 &&x=\frac{a}{a_{ta}},y=\frac{a_p}{a_{p,ta}}, \nu=\frac{\rho_{Qb,ta}}{\rho_{mb,ta}}
 \label{xynuDefinition}
 \eeq
 and using the fact that $[\dot{a}_p]_{ta}=0$ we can get
 \beq{}
 (\frac{dy}{dx})^2=\frac{\zeta y^{-1}+\nu\zeta^{1+w}y^{-1-3w}-(\zeta+\nu\zeta^{1+w})}{x^{-1}+\nu
 x^{-1-3w}}.\label{dydxMainEqQCDM}
 \eeq
 Then using eq(\ref{roaagoesto0}) and
 the approximate mass conserving condition we can prove
 \beq{}
 &&[\frac{y}{x}]_{x\rightarrow0}=[\frac{a_pa^{-1}_{p,ta}}{aa^{-1}_{ta}}]_{a\rightarrow0}=1_{-}\cdot\zeta^{\frac{1}{3}}
 \label{}
 \eeq
 and
 \beq{}
 \frac{d}{dx}[y_{x\rightarrow0,x\neq0}]=\zeta^{\frac{1}{3}}(1-2\alpha\cdot a_{ta}x).
 \label{dydx}
 \eeq
 Substituting eqs(\ref{roaagoesto0}), (\ref{xynuDefinition}) and (\ref{dydx}) into
 eq(\ref{dydxMainEqQCDM}), taking limit $x\rightarrow0$ and preserving only the
 first order small quantities in $x$, we can get
 \beq{}
 \alpha=\frac{1}{5}a^{-1}_{ta}[\zeta^{\frac{1}{3}}+\nu\cdot
 \zeta^{\frac{1}{3}+w}]
 \eeq
 Substituting this result into eq(\ref{deltacOperationDefinition})
 we can get
 \beq{}
 \delta_c(w,\Omega_{m0},a_c)=\frac{3}{5}a_{ta}^{-1}[\zeta^{\frac{1}{3}}+\nu\cdot
 \zeta^{\frac{1}{3}+w}]D_1(a_c)
 \label{QCDMdeltacAnalytical}
 \eeq

 Now let us come to $\Delta_c^\prime$s calculation.
 According to the definition eq(\ref{DeltacDefinition1}),
 \beq{}
 \Delta_c(w,\Omega_{m0},a_c)=\frac{\rho_{mc,c}}{\rho_{mb,c}}
 =\frac{\rho_{mc,ta}a_{p,ta}^3a_{p,c}^{-3}}{\rho_{mb,ta}a_{ta}^3a_c^{-3}}
 =\zeta\frac{a_{p,ta}^3}{a_{p,c}^{3}}\frac{a_c^3}{a_{ta}^3}
 \nonumber\\
 \label{DeltacDefinition2}
 \eeq
 To calculate the second factor of the above equations' right-most
 part, we can use energy conserving condition and virial theorem.
 Assuming that at the collapse point, the system virialized fully,
 then we have the following relations:
 \beq{}
 &&E_{kinetic,c}=-\frac{1}{2}U_{mm,c}+U_{mQ,c}-\frac{1}{2}U_{QQ,c}
 \label{virialTheoremQCDM}\\
 &&\frac{1}{2}U_{mm,c}+2U_{mQ,c}+\frac{1}{2}U_{QQ,c}\nonumber\\
 &&\hspace{15mm}=U_{mm,ta}+U_{mQ,ta}+U_{QQ,ta}
 \label{EConsQCDM}
 \eeq
 where
 $U_{mm}$, $U_{mQ}$ and $U_{QQ}$ denote the matter-matter,
 matter-Quintessence and Quintessence-Quintessence gravitation
 potentials respectively. The subscripts $_{,c}$ and $_{,ta}$ mean that the
 quantities carrying them take values at the collapse and turn
 around time respectively. Explicitly, eq(\ref{EConsQCDM}) can be
 written out as:
 \beq{}
 &&\hspace{-3mm}(\rho_{mc,c}^2+4\rho_{mc,c}\rho_{Qc,c}+\rho_{Qc,c}^2)a_{p,c}^5
 \nonumber\\
 &&=2(\rho_{mc,ta}^2+\rho_{mc,ta}\rho_{Qc,ta}+\rho_{Qc,ta}^2)a_{p,ta}^5
 \eeq
 dividing each side of this equation by the relation
 $\rho_{mc,c}^2a_{p,c}^6=$ $\rho_{mc,ta}^2a_{p,ta}^6$
 and making a little deformation we get:
 \beq{}
 [\frac{a_{p,ta}}{a_{p,c}}]
 (1+4\frac{\rho_{Qc,c}}{\rho_{mc,c}}+\frac{\rho_{Qc,c}^2}{\rho_{mc,c}^2})
 =2(1+\frac{\rho_{Qc,ta}}{\rho_{mc,ta}}+\frac{\rho_{Qc,ta}^2}{\rho_{mc,ta}^2})
 \nonumber\\
 \label{aptaOapc_def_QCDM1}
 \eeq
 Using energy conservation law and the approximate mass
 conserving condition, we can change eq(\ref{aptaOapc_def_QCDM1})
 into the following form:
 \beq{}
 [\frac{a_{p,ta}}{a_{p,c}}]
 (1+4\xi [\frac{a_{p,ta}}{a_{p,c}}]^{3w}+\xi^2 [\frac{a_{p,ta}}{a_{p,c}}]^{6w})
 =2(1+\xi+\xi^2)\nonumber\\
 \label{aptaOapc_def_QCDM2}
 \eeq
 where
 \beq{}
 \xi=\nu\zeta^{w}
 \eeq
 Numerical studying shows that the third term on both sides of
 eq(\ref{aptaOapc_def_QCDM2}) can be neglected, just as
 \cite{WangSteinhardt1} did. But, for a general
 valued $w$, even neglecting this two terms, this equation cannot be
 solved analytically. So we preserve this two terms in the
 equation and resort to numerical methods to calculate
 $\frac{a_{p,ta}}{a_{p,c}}$.

 From eqs(\ref{QCDMdeltacAnalytical}) and
 (\ref{DeltacDefinition2}),(\ref{aptaOapc_def_QCDM2}) we see that,
 given $w$, $\Omega_{m0}$ and $a_c$, to calculate
 $\delta_c$ and $\Delta_c$, we need to express $a_{ta}$, $\nu$ and
 $\zeta$ as functions of $w$, $\Omega_{m0}$ and $a_c$.
 For $\nu$
 \beq{}
 &&\nu=\frac{1-\Omega_{mb,ta}}{\Omega_{mb,ta}}
 \label{nuofOm0andac}\\
 &&\Omega_{mb,ta}=\frac{1}{1+\nu_0a_{ta}^{-3w}},\
 \nu_0=\frac{1-\Omega_{m0}}{\Omega_{m0}}
 \eeq
 While for $a_{ta}$, we can use the fact that
 $t_c=2t_{ta}$ and eq(\ref{FriedmanBackground})
 to set up an integration equation
 \beq{}
 \int_0^{a_c}da^{\prime}\sqrt{\frac{a^{\prime}}{1+\nu_0 a^{\prime
 -3w}}}=2\int_0^{a_{ta}}da^{\prime}\sqrt{\frac{a^{\prime}}{1+\nu_0 a^{\prime -3w}}}
 \nonumber\\
 \label{QCDMtcandtta}
 \eeq
 In the case of
 $w=-1$, eq(\ref{QCDMtcandtta}) can be solved analytically:
 \beq{}
 a_{ta}=\left[\frac{\sqrt{1+\nu_0a^3_c}-1}{2\nu_0}\right]^{1/3}
 \label{LCDMtcAndtta}
 \eeq
 While in the case of general valued $w$, eq(\ref{QCDMtcandtta}) should
 be solved numerically to get the function $a_{ta}(a_c)$.

 For $\zeta$'s calculation, there is two method. The simplest is
 directly integrate eq(\ref{dydxMainEqQCDM}) to set up an
 equation:
 \beq{}
 &&\int_0^1dx[\frac{x}{1+\nu x^{-3w}}]^{\frac{1}{2}}=
 \nonumber\\
 &&\hspace{5mm}\int_0^1dy[\frac{y}{\zeta+\nu\zeta^{1+w}y^{-3w}-(\zeta+\nu\zeta^{1+w})y}]^{\frac{1}{2}}
 \label{zetaIntEq}
 \eeq
 Solving this equation will tell us $\zeta$'s
 dependence on $w$ and $\nu$. Combining the results with
 eq(\ref{nuofOm0andac}) will gives us $\zeta$'s dependence on $w$,
 $\Omega_{m0}$ and $a_c$.
 The second method is to use eqs(\ref{FriedmanBackground}),
 (\ref{time_time_Einstein}) and the notations in
 eq(\ref{zetaDefinition})-(\ref{xynuDefinition}) to set up an eigen-value problem
 \beq{}
 &&[\frac{y}{x}]_{x\rightarrow0}=1_{-}\cdot\zeta^{\frac{1}{3}}
 \nonumber\\
 &&y|_{x=1}=1,\ y^{\prime}|_{x=1}=0.
 \label{BC-zeta-eigenValueEqQCDM}\\
 &&\hspace{-3mm}\frac{d^2y}{dx^2}
 -\frac{dy}{dx}\frac{1}{2}[\frac{1}{x}+\frac{d\Omega_{mb}}{dx}\frac{1}{\Omega_{mb}}]
 \nonumber\\
 &&\hspace{-2mm}+\frac{1}{2}[(1+3w)[\frac{y}{x}]^{-3(1+w)}\zeta^{1+w}\frac{y(1-\Omega_{mb})}{x^2}+\frac{\zeta
 x\Omega_{mb}}{y^2}]=0\nonumber\\
 \label{zetaDeterminingEqQCDM}
 \eeq
 Solving this eigen-value problem can also tell us $\zeta$'s
 dependence on $w$ and $\nu$ \cite{PressNR}. For a given value of $w$, we can
 get $\zeta$'s dependence on $\Omega_{mb,ta}$ by solving
 eq(\ref{zetaIntEq}) or
 (\ref{BC-zeta-eigenValueEqQCDM})-(\ref{zetaDeterminingEqQCDM})
 and fitting the results as $\zeta=\zeta(\Omega_{mb,ta})$. For
 example, in the case of $w=-1$,
 \beq{}
 \zeta=[\frac{3\pi}{4}]^2\Omega_{mb,ta}^{-0.7384+0.2451\Omega_{mb,ta}}
 \label{LCDMzetaFit}
 \eeq
 But for the general cases when $\Omega_{mb,ta}$ and $w$ both vary,
 we cannot fit the results into a simple function, see
 Fig.\ref{zeta}. It should be noted that this kind of fitting
 formula eq(\ref{LCDMzetaFit}) only
 coincide with numerical result when $\Omega_{mb,ta}$ is not too
 small, for example, $\Omega_{mb,ta}>0.05$.

 \subsection{Numerical Results, Effects of
 $w$ on the Key Parameters of the Model}

\begin{figure}
 \includegraphics[scale=0.39]{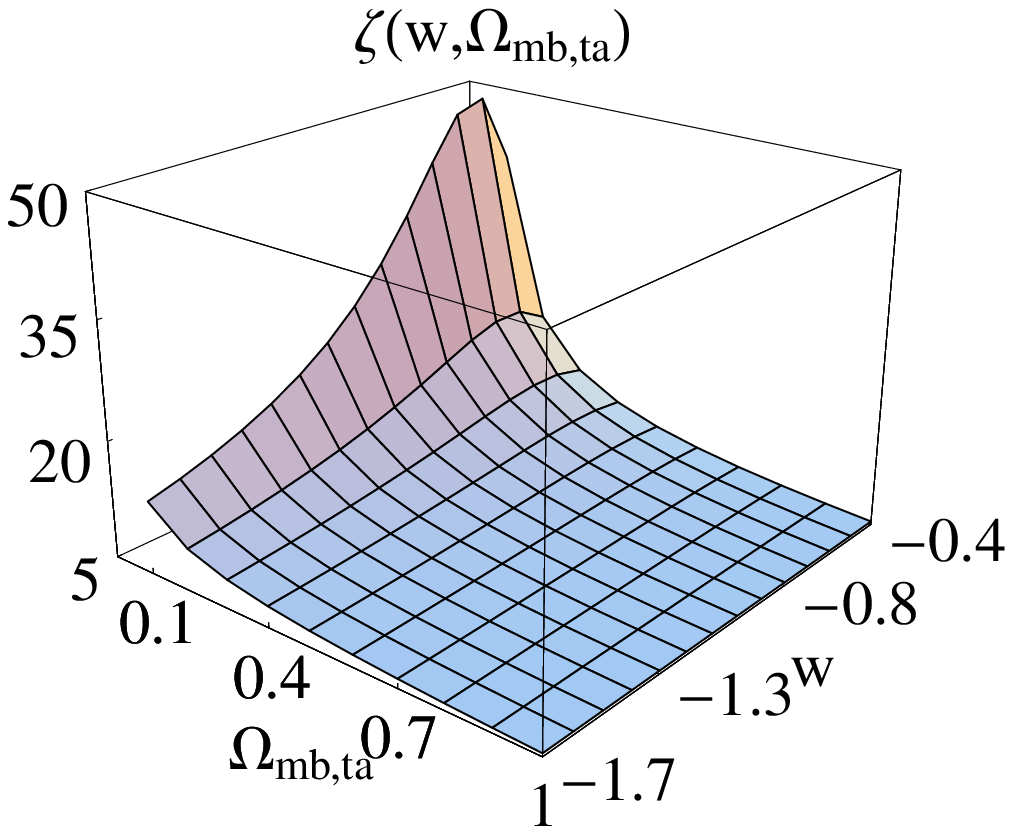}
 \includegraphics[scale=0.39]{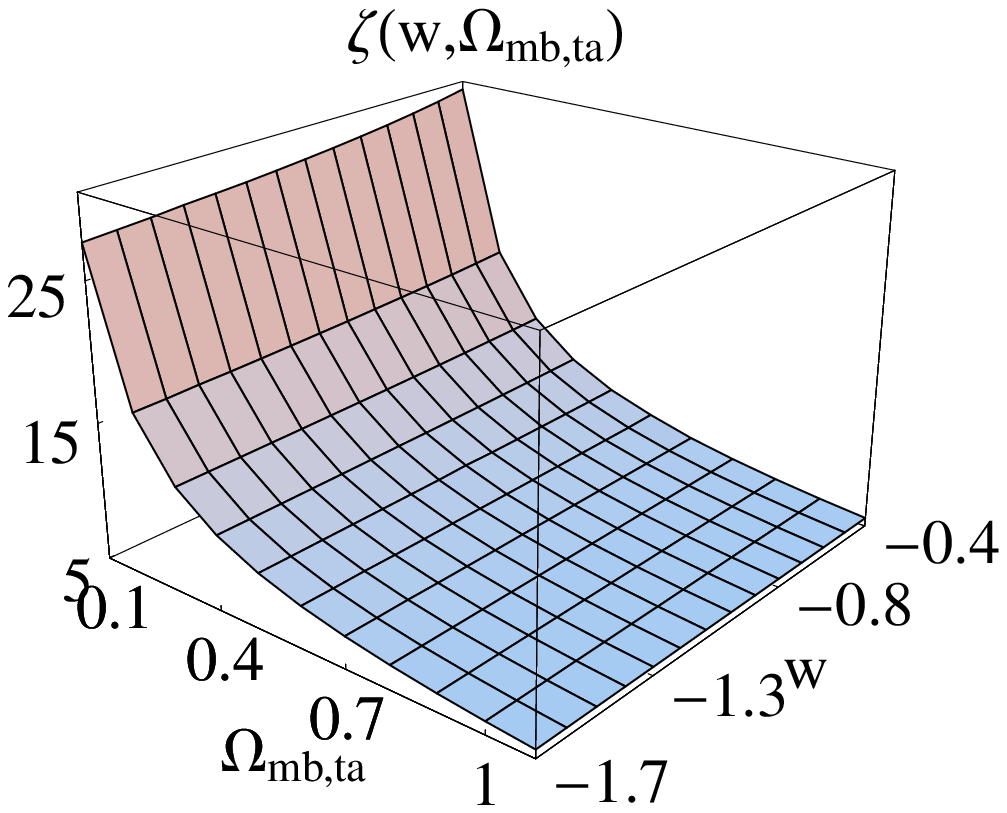}
 \caption{$\zeta$'s dependence on $\Omega_{mb,ta}$ and $w$. Left
 panel, considering Quintessence's clustering effects. It is solved from
 eq(\ref{zetaIntEq}) or (\ref{zetaDeterminingEqQCDM}). Right
 panel, the results of assuming Quintessence does not cluster
 on the scale of galaxy clusters but neglect the Q-current
 flowing outside the clusters, solved from
 eq(A9) of \cite{WangSteinhardt1}.
 }
 \label{zeta}
 \end{figure}

\begin{figure}
 \includegraphics[scale=0.4]{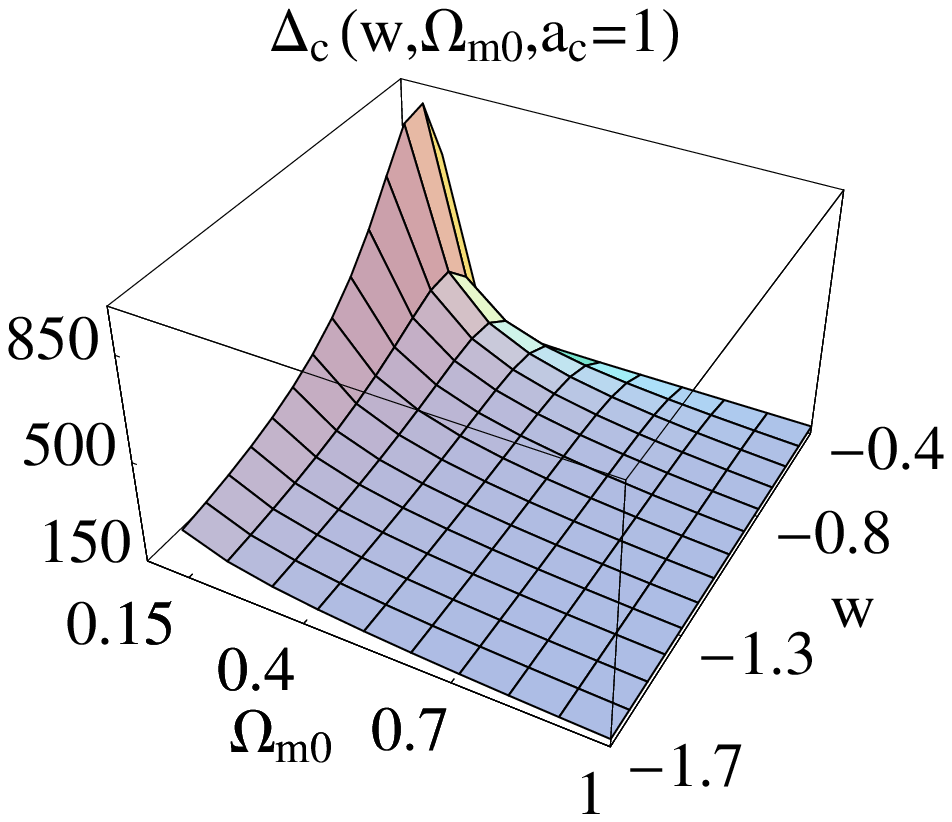}
 \includegraphics[scale=0.4]{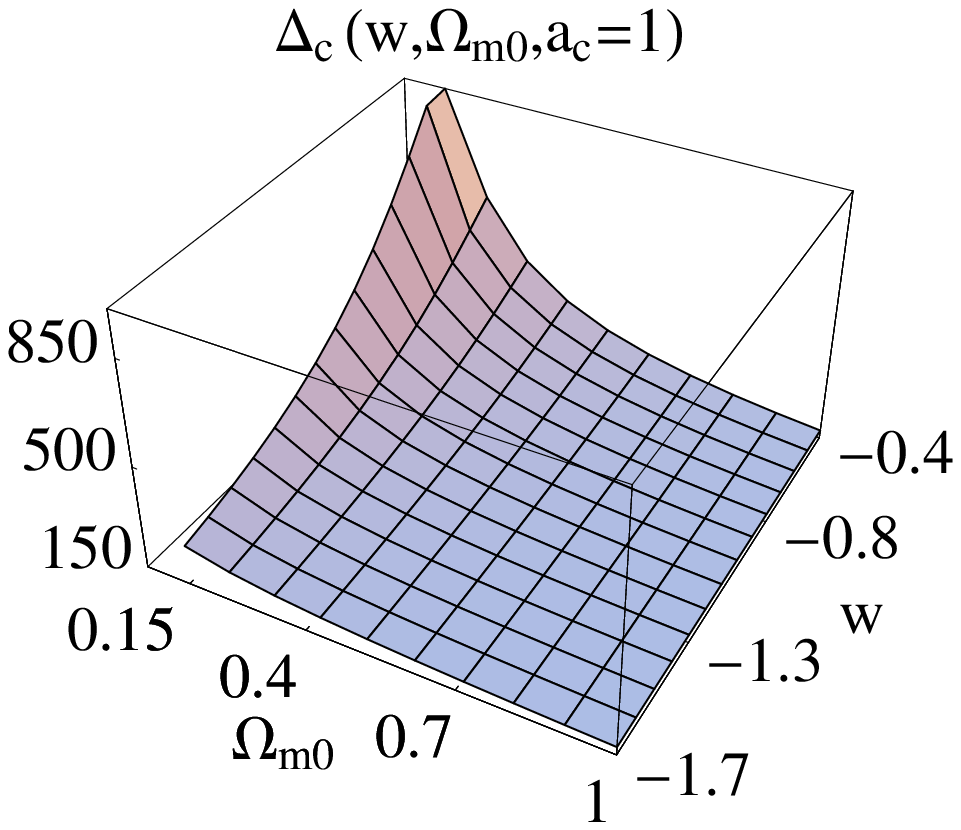}
 \\
 \includegraphics[scale=0.4]{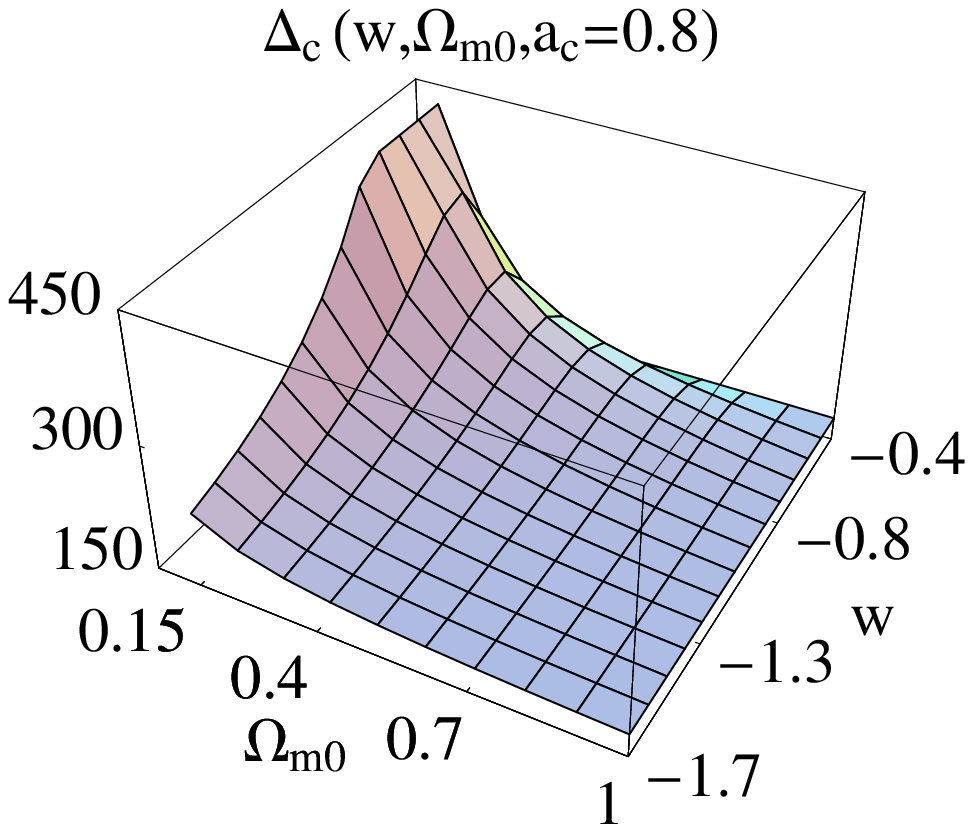}
 \includegraphics[scale=0.4]{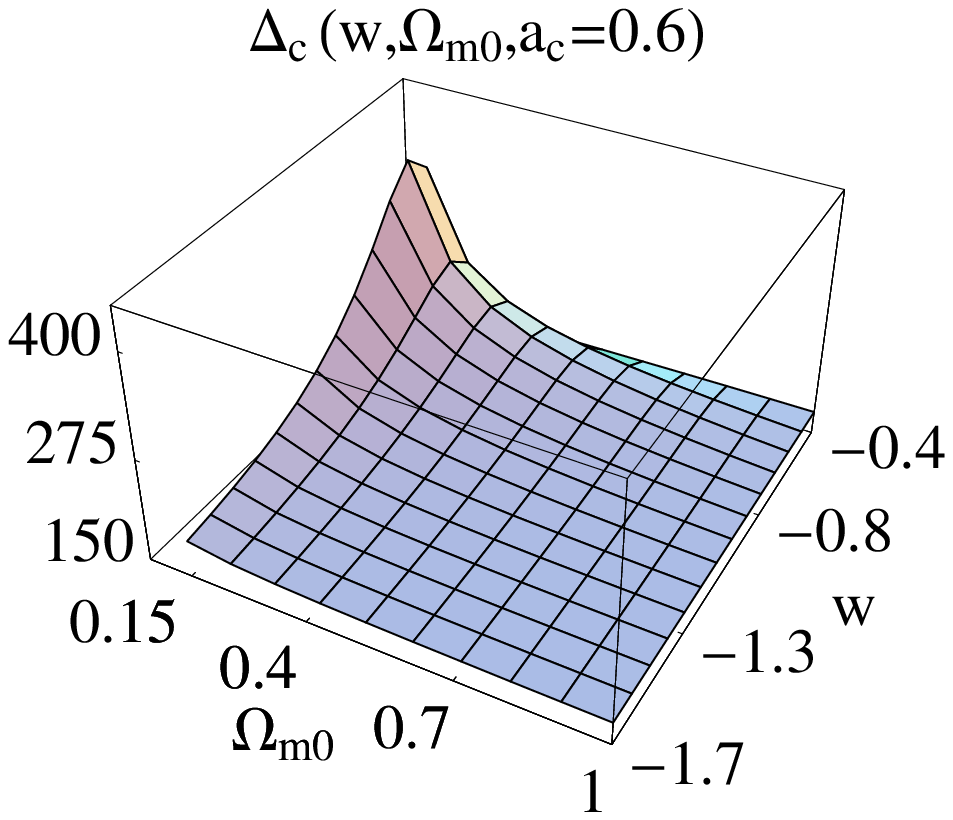}
 \caption{$\Delta_c$'s dependence on $\Omega_{m0}$, $w$ and $a_c$.
 The upper right part is
 solved from eq(5)-(6) of \cite{WangSteinhardt1} where
 Quintessence is assumed not to cluster on the scale of galaxy clusters
 but neglect the Q-current flowing outside the over-dense region.
 The others are
 solved from eq(\ref{zetaIntEq}) or (\ref{zetaDeterminingEqQCDM})
 of this paper.
 }\label{Dccompare}
 \end{figure}

\begin{figure}
 \begin{minipage}[c]{0.25\textwidth}
 \includegraphics[scale=0.43]{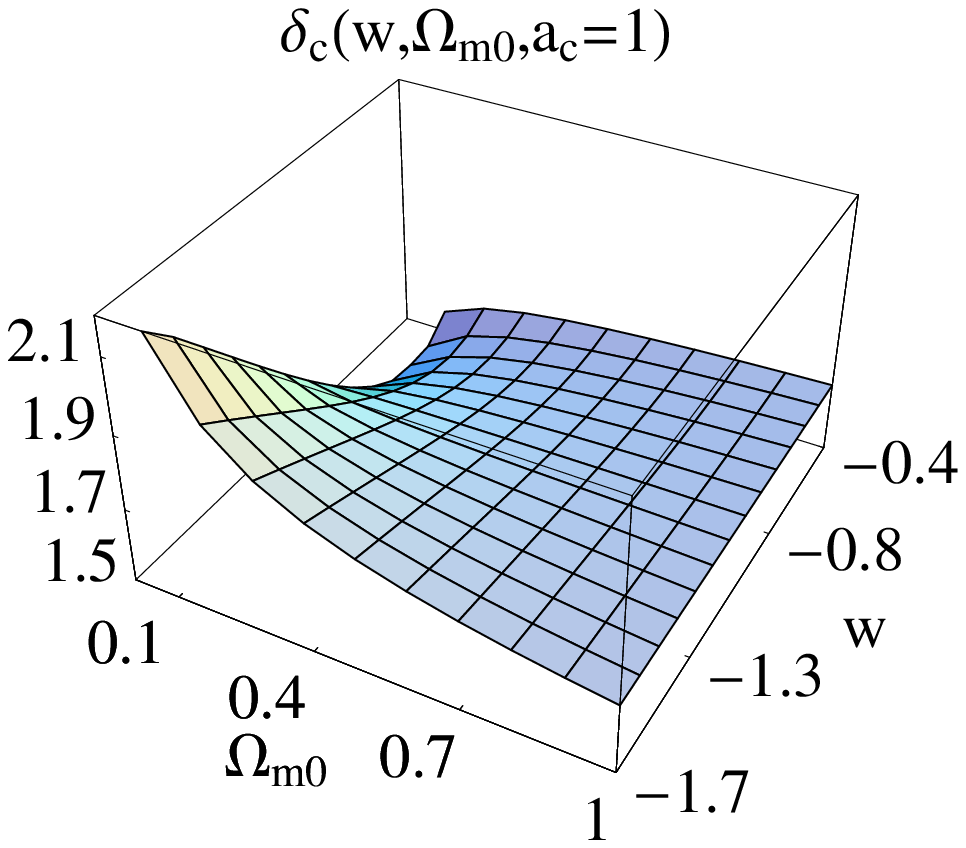}
 \end{minipage}%
 \begin{minipage}[c]{0.25\textwidth}
 \includegraphics[scale=0.4]{fig3b.eps}
 \end{minipage}
 \\
 \includegraphics[scale=0.4]{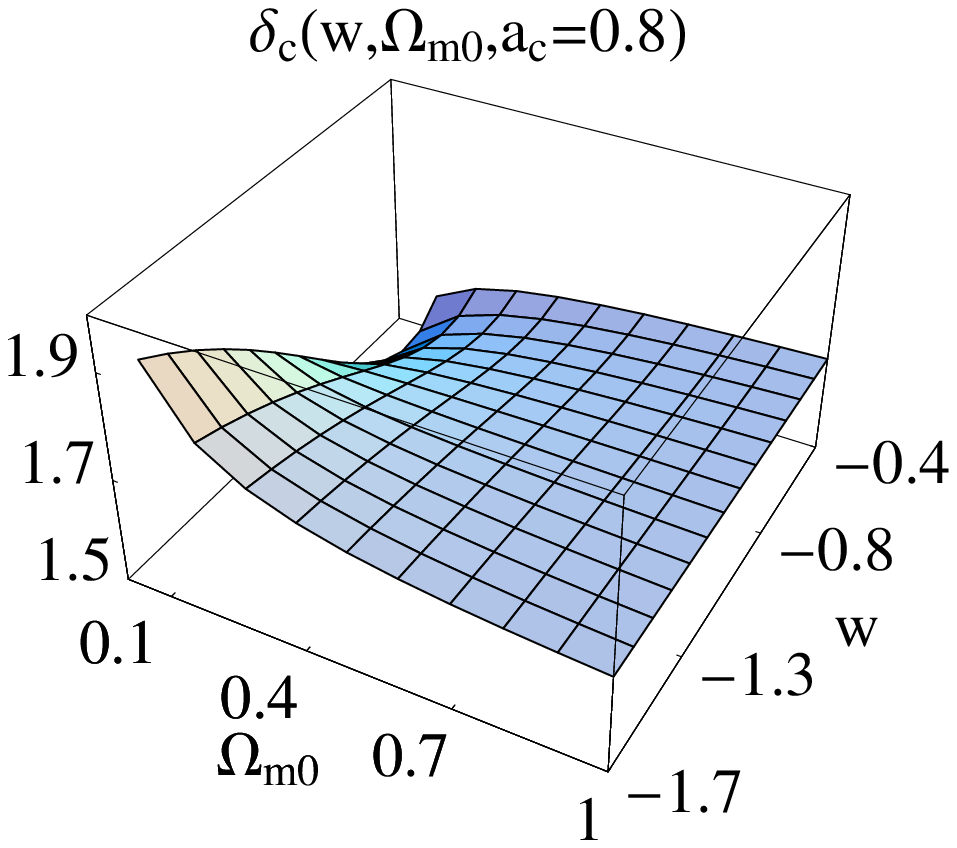}
 \includegraphics[scale=0.4]{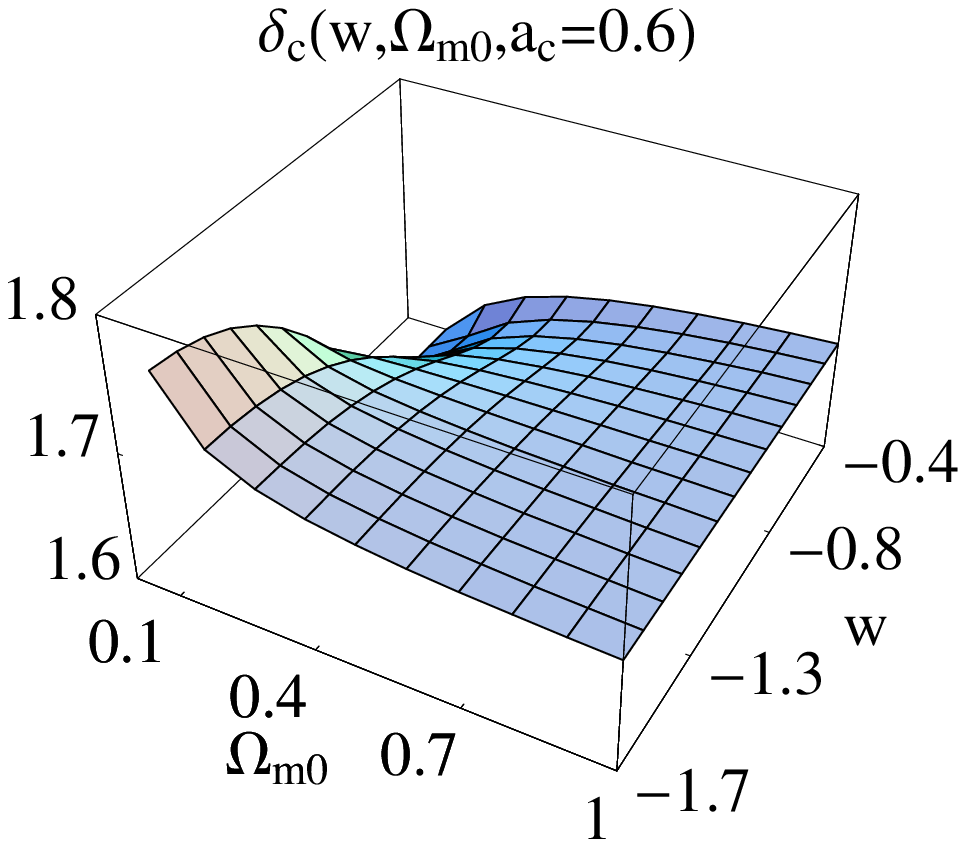}
 \caption{$\delta_c$'s dependence on $w$, $\Omega_{m0}$ and $a_c$.
 Quintessence's clustering effects are considered.
 }\label{dc}
 \end{figure}

 We provide our numerical results in FIG.\ref{zeta}-\ref{dc}.
 From FIG.\ref{zeta} we see that, when the dark energy's clustering
 is considered, $\zeta$ depends on the value of $w$ remarkably.
 For a fixed $\Omega_{mb,ta}$, $\zeta$ first increases then
 decreases as $w$ decreases in the range $[-1.7,-0.4]$,
 this forms a strong contrast with the result of \cite{WangSteinhardt1},
 which is also depicted in the figure.
 In FIG.\ref{Dccompare}, besides similar
 dependence of $\Delta_c$ on $w$ and $\Omega_{m0}$, we can
 still see that for a given $\Omega_{m0}$ and $w$,
 $\Delta_c$ decreases as $a_c$ decreases, i.e. the early an
 over-dense region virialized, the less is its matter density over
 that of the background. This is easy to understand, because the
 earlier an over-dense region virializes, the less dark energy is
 contained in the background cosmology, so the weaker the
 anti-cluster effects will be and the less dense a region is
 required to keep balance by self-gravitations.


 In FIG.\ref{dc}, we display $\delta_c$'s dependence on
 $w$, $\Omega_{m0}$ and $a_c$. Unlike in FIG.\ref{Dccompare},
 we did not compare $\delta_c$'s
 dependence on $w$, $\Omega_{m0}$ in this figure between the cases
 where Quintessence is assumed to cluster
 and the contrary.
 By the method of \cite{WangSteinhardt1}, eq(A12)-(A13),
 it is difficult to get the precise value of
 $\delta_c(w,\Omega_{m0},a_c)$. Since this method
 needs to take the value of
 $[\frac{x}{y}]$ at some point $x=x_0\rightarrow0,x_0\neq0$,
 while the coefficients of the differential
 equation which is used to solve $\zeta$ has singularities
 near this region, the quantity $\delta_c(x)$ calculated by this method
 depends on the selection of
 $x_0$ rather artificially. By the way, we note that eq(A12) may missed a term
 $\zeta$ on the right hand side. By the statement of \cite{WangSteinhardt1},
 $\delta_c\approx1.6-1.686$, where $-1<w<0$. Here, in our
 FIG.\ref{dc}, we see that, if $a_c=1$, $\delta_c$ may be as large
 as $2.1$ when $w\approx-1.7$. As $a_c$ decreases, $\delta_c$ also
 decreases and goes to the limit value 1.686, what ever value $w$
 and $\Omega_{m0}$ takes. This is trivial, because the earlier an
 over-dense region virializes, the more likely is the back ground
 universe a totaly matter dominated one.

 \section{The Number Density of Galaxy Clusters and Its Evolutions
 }\label{PS_App_Section}

 \subsection{Theoretical Formulaes}\label{theoreticalFormulaSec}

 According to Press-Schechter theory, the comoving number density
 of clusters which have collapsed (i.e., virialized) at certain
 red-shift $z$ and
 have masses in the range $M\sim M+dM$ could be calculated:
 \beq{}
 n(M,z)dM&&\hspace{-3mm}=
 -\sqrt{\frac{2}{\pi}}\frac{\rho_{tot}}{M}
 \frac{\delta_c}{\sigma^2}
 \frac{R}{3M}\frac{d\sigma}{dR}
 \textrm{exp}[-\frac{\delta_c^2}{2\sigma^2}]dM
 \label{diff-M-Function2}
 \eeq
 in this equation: $\rho_{tot}$ is the energy density of background universe,
 $M=\frac{4\pi}{3}R^3\rho_{tot}$, so the factor
 $\frac{\rho_{tot}}{M}$ denotes the average number density of
 clusters with mass $M$. The other factors in this equations are
 just obtained by differentiating eq(\ref{Press-Schechter}) with
 respect to $M$. In the original paper of Press and Schechter
 \cite{PressSchechter}, the quantity $\rho$ and $M$ appearing in
 eq(\ref{diff-M-Function2}) only refer to density and mass of
 matters. But under our assumptions, Quintessence
 and matters cluster synchronously so we have to understand
 them as the total density and mass.

 To relate the mass of a cluster
 with its characteristic X-ray temperature,
 consider a virialized over-dense spherical region
 in the background universe containing quintessences,
 according to virial theorem, we have
 \beq{}
 E_{\textrm{kinetic,vir}}=\left[-\frac{1}{2}U_{mm}+U_{mQ}-\frac{1}{2}U_{QQ}\right]_{,vir}
 \label{VirialTheorem}
 \eeq
 i. e.
 \beq{}
 (\rho_{mc}+\rho_{Qc})_{,vir}\bar{V}_{vir}^2=\frac{4\pi
 G}{5}\left[a_{p}^2(\rho_{mc}-\rho_{Qc})^2\right]_{,vir}
 \nonumber
 \eeq
 where $\bar{V}^2_{vir}$ is the mean
 square velocity of particles in the cluster when the system is
 fully virialized and $a_p$ is the physical radius of the cluster.
 From this equation, we have
 \beq{}
 \bar{V}^2_{vir}=\frac{3}{5}\left[(GMH)^\frac{2}{3}
 [\frac{1}{2}\frac{\rho_{mc}+\rho_{Qc}}{\rho_{mb}+\rho_{Qb}}]^{\frac{1}{3}}(1-\frac{2\rho_{Qc}}{\rho_{mc}+\rho_{Qc}})^2\right]_{,vir}
 \nonumber
 \eeq
 Using relation:
 \beq{}
 k_BT=\frac{\mu
 m_p}{\beta}\frac{\bar{V}^2_{vir}}{3}
 \label{vTrelation}
 \eeq
 where $k_B$ is
 the Boltzmann constant, $m_p$ is the mass of proton, while $\mu m_p$ is
 the average mass of particles, $\beta$ is the ratio of kinetic energy to
 temperature. So we have mass-temperature relation:
 \beq{}
 M=\frac{1}{GH(z)}\left[\frac{5\beta k_BT}{\mu m_p}\frac{1}{f(z)}\right]^{\frac{3}{2}}
 \eeq
 or
 \beq{}
 R&&\hspace{-3mm}=\left[\frac{2GM}{H^2}\right]^{\frac{1}{3}}
 =\frac{1}{H(z)}\left[\frac{5\cdot2^{\frac{2}{3}}\beta k_BT}{\mu m_p}\frac{1}{f(z)}\right]^{\frac{1}{2}}
 \label{RTrelation}
 \eeq
 with
 \beq{}
 f(z)&&\hspace{-3mm}=[\frac{1}{2}(\Omega_{mb,c}+\Omega_{Qb,c}\Delta_c^w)\Delta_c]^{\frac{1}{3}}\times
 \nonumber\\
 &&\hspace{10mm}(1-\frac{2\Omega_{Qb,c}\Delta_c^w}{\Omega_{mb,c}+\Omega_{Qb,c}\Delta_c^w})^2
 \nonumber\\
 H(z)&&\hspace{-3mm}=H_0[\Omega_{m0}(1+z)^{3}+(1-\Omega_{m0})(1+z)^{3(1+w)})]^{\frac{1}{2}}
 \nonumber\\
 \eeq
 and $\Delta_c$ given by eq(\ref{DeltacDefinition2}).

 Just as \cite{WangSteinhardt1} pointed out, since the
 mass-temperature relation is red-shift dependent, simply
 substituting eq(\ref{RTrelation}) into eq(\ref{diff-M-Function2})
 cannot give us correct number density of clusters in a given
 temperature range today. Instead, we should first find out
 the virialization rate and multiply it by the mass-temperature relation
 then integrate over red-shift
 \beq{}
 &&\hspace{-3mm}n(T,z)dT=\nonumber\\
 &&-\frac{1}{\sqrt{2\pi}}\int_z^{\infty}\frac{\rho_{tot}}{MT}
 \frac{d\textrm{ln}\sigma}{d\textrm{ln}R}
 \frac{d\textrm{ln}\sigma}{dz}
 \frac{\delta_c}{\sigma}(\frac{\delta_c^2}{\sigma^2}-1)
 \textrm{exp}[-\frac{\delta_c^2}{2\sigma^2}]dzdT
 \nonumber\\
 \label{diff-T-Function}
 \eeq

 From eqs(\ref{diff-T-Function}), (\ref{RTrelation}),
 (\ref{QCDMdeltacAnalytical}) and (\ref{sigmaRz}) we can see that
 besides the constant $\frac{\mu}{\beta}$,
 $n(T,z)$ will also depend on the cosmological parameters
 $w$, $\Omega_{m0}$, $h$, $n_s$ and the normalization $\sigma_8$
 of the cosmic density fluctuations. In principle, if we can
 measure the number density v.s. temperature relation precisely
 enough, by numerical fittings, we can determine all these parameters
 or some of their special combinations simultaneously from observations.

 \begin{figure}
 \includegraphics[scale=0.21]{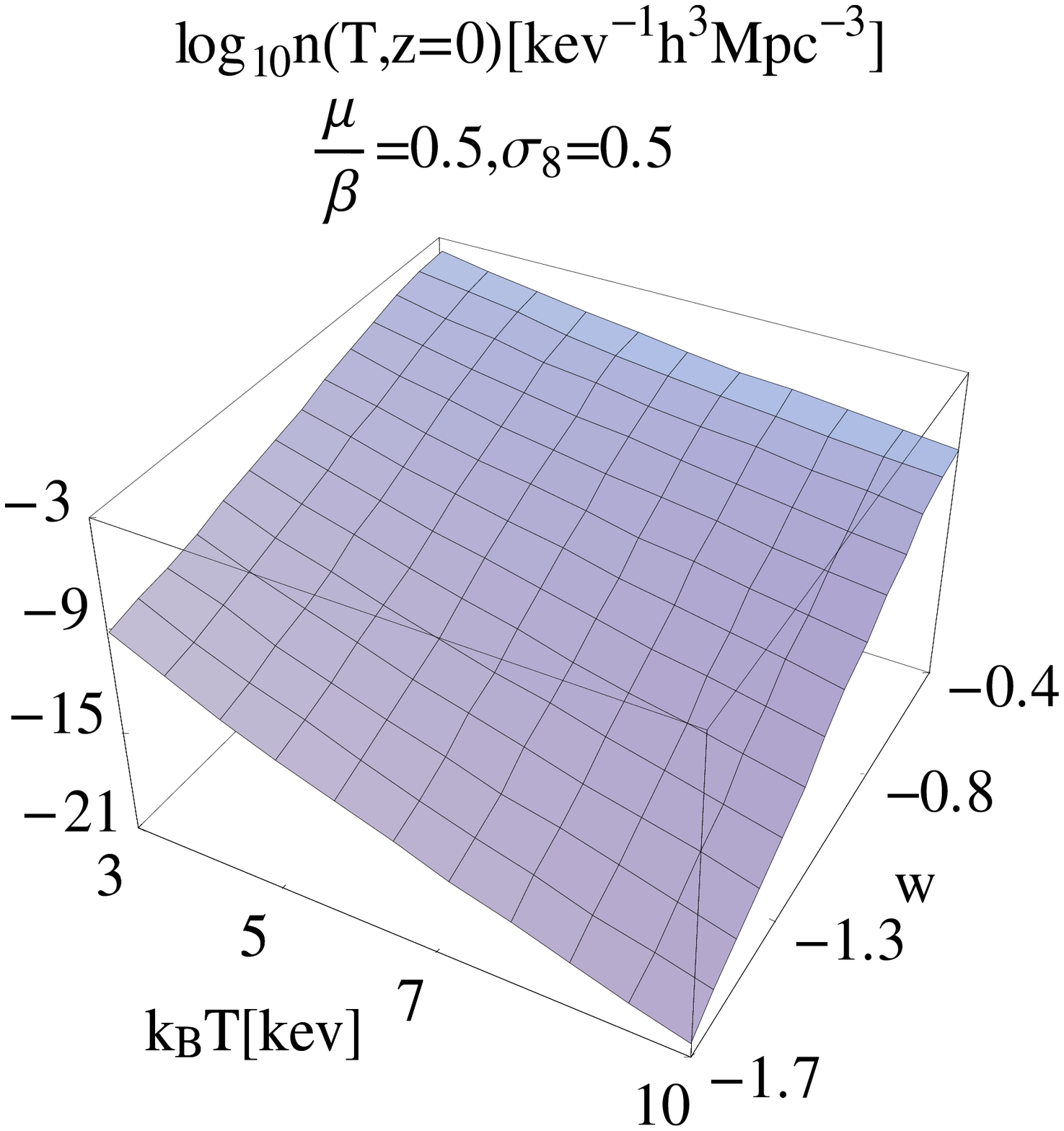}
 \includegraphics[scale=0.21]{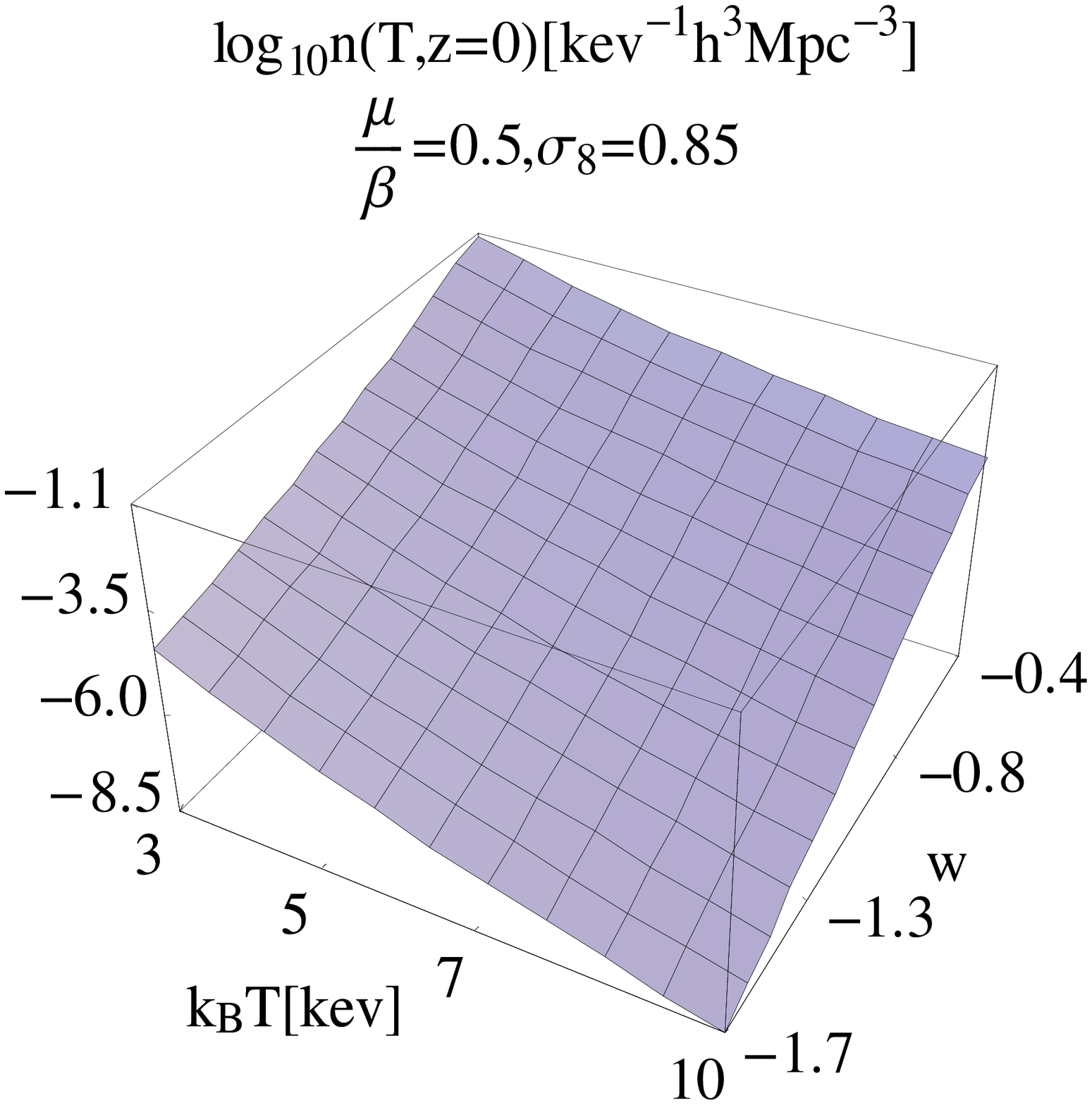}\\
 \includegraphics[scale=0.4]{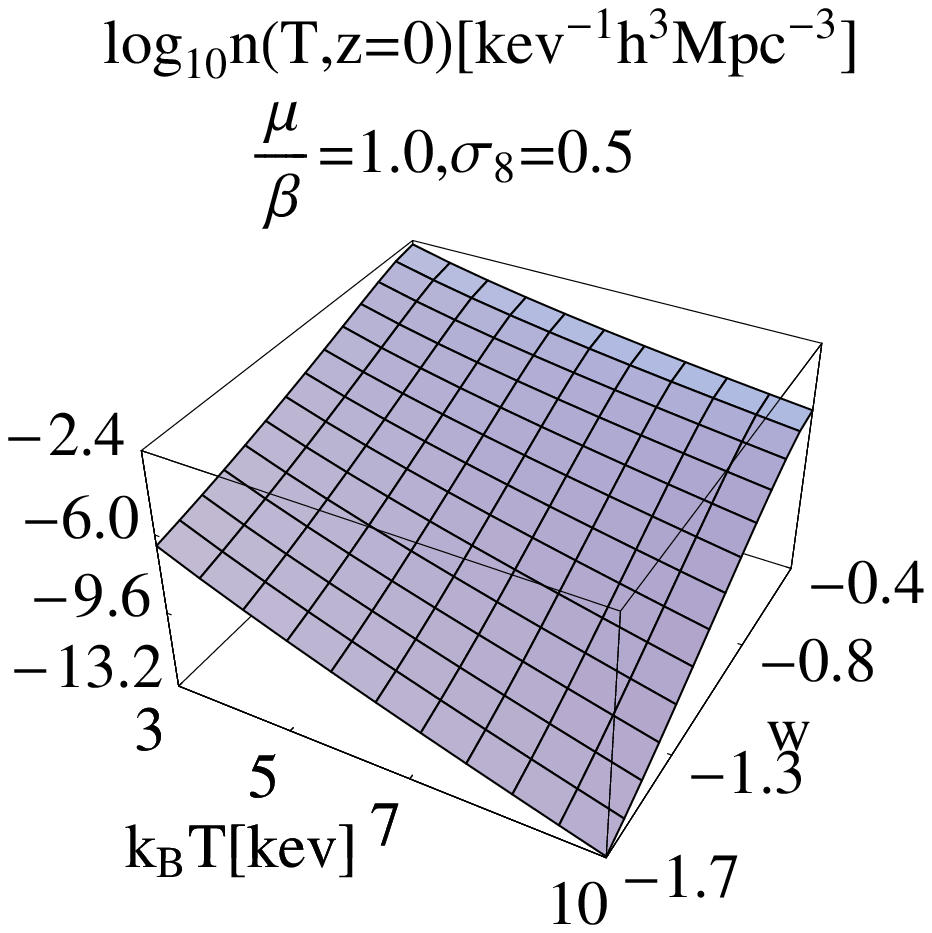}
 \includegraphics[scale=0.4]{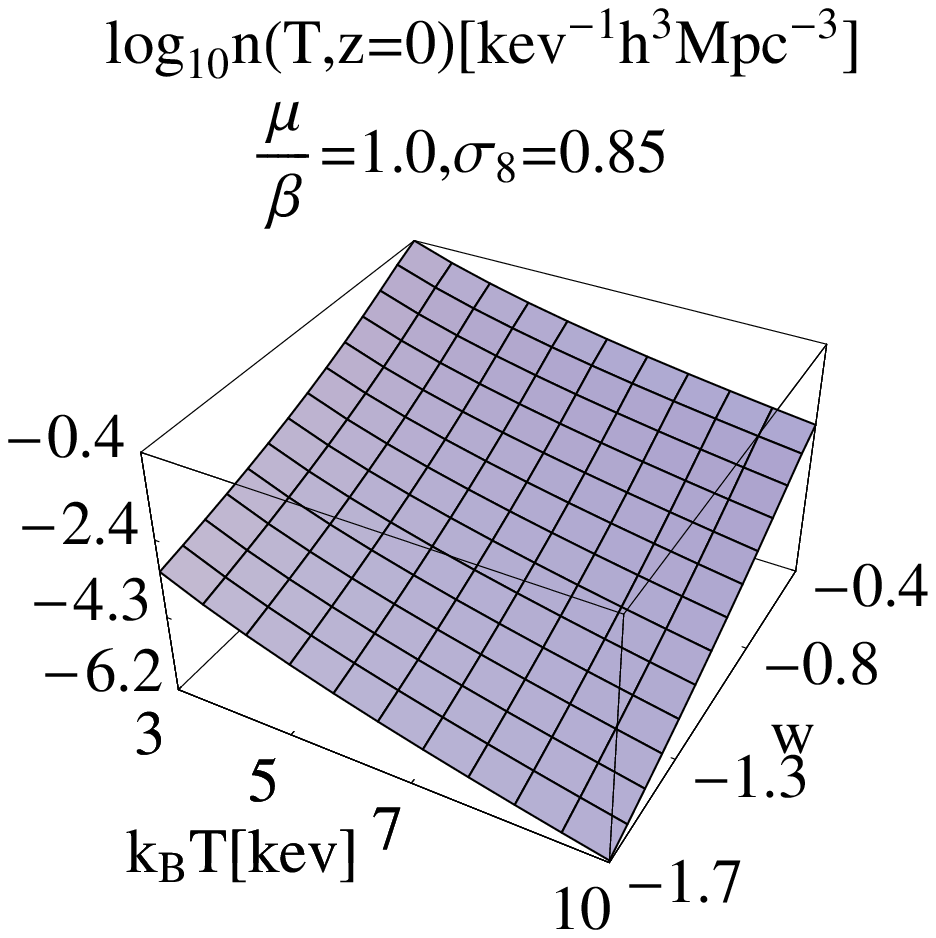}
 \caption{Effects of $w$ on the
 number density v.s. temperature function  of galaxy clusters when
 $z=0$. The larger is $\sigma_8$ or $\frac{\mu}{\beta}$, the
 larger the function value will be.
 All four figures have
 $\Omega_{m0}=0.27$, $h=0.71$ and $n_s=1.0$.
 }
 \label{ew_nTcore}
 \end{figure}

 \begin{figure}
 \includegraphics[scale=0.21]{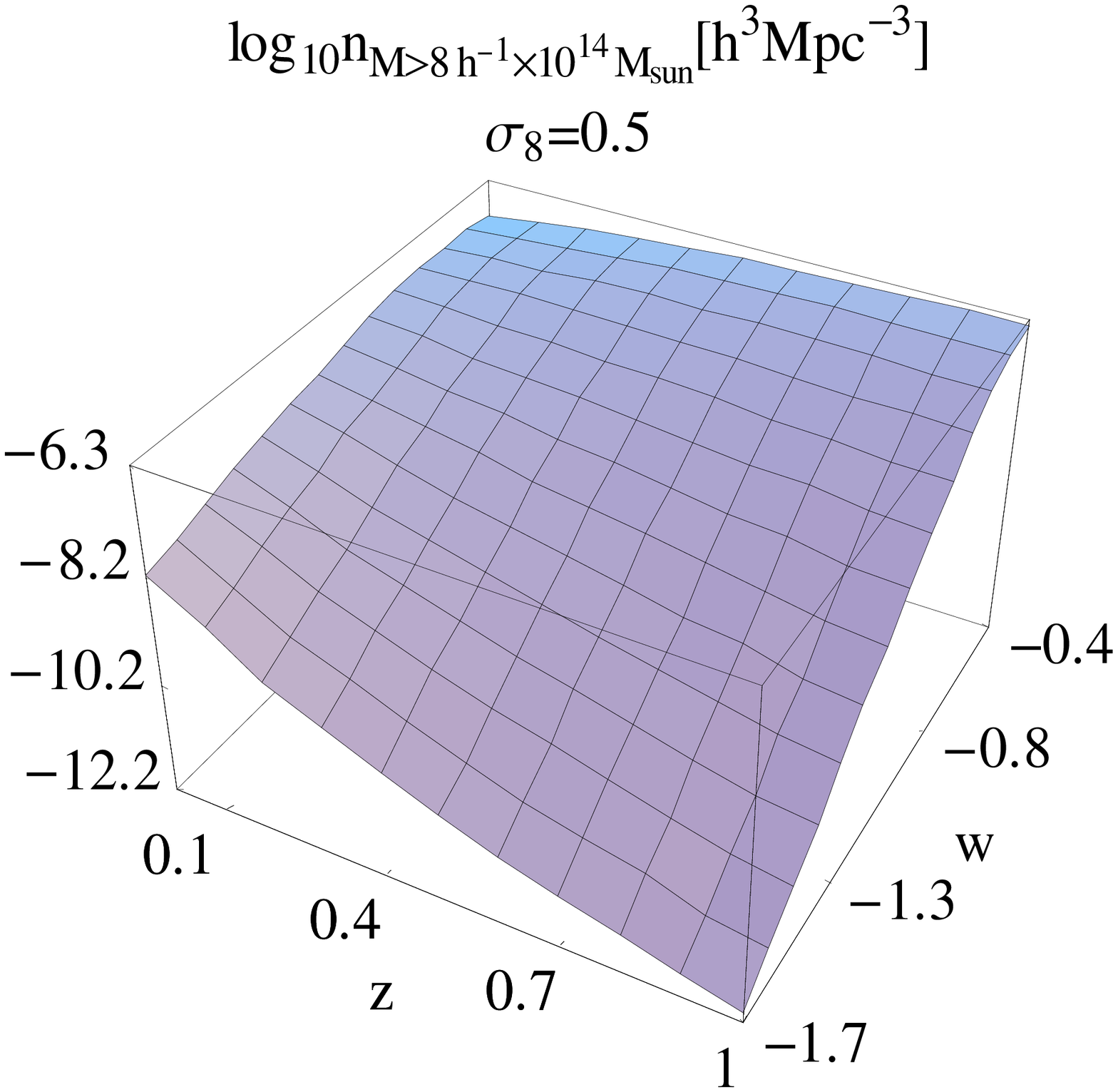}
 \includegraphics[scale=0.21]{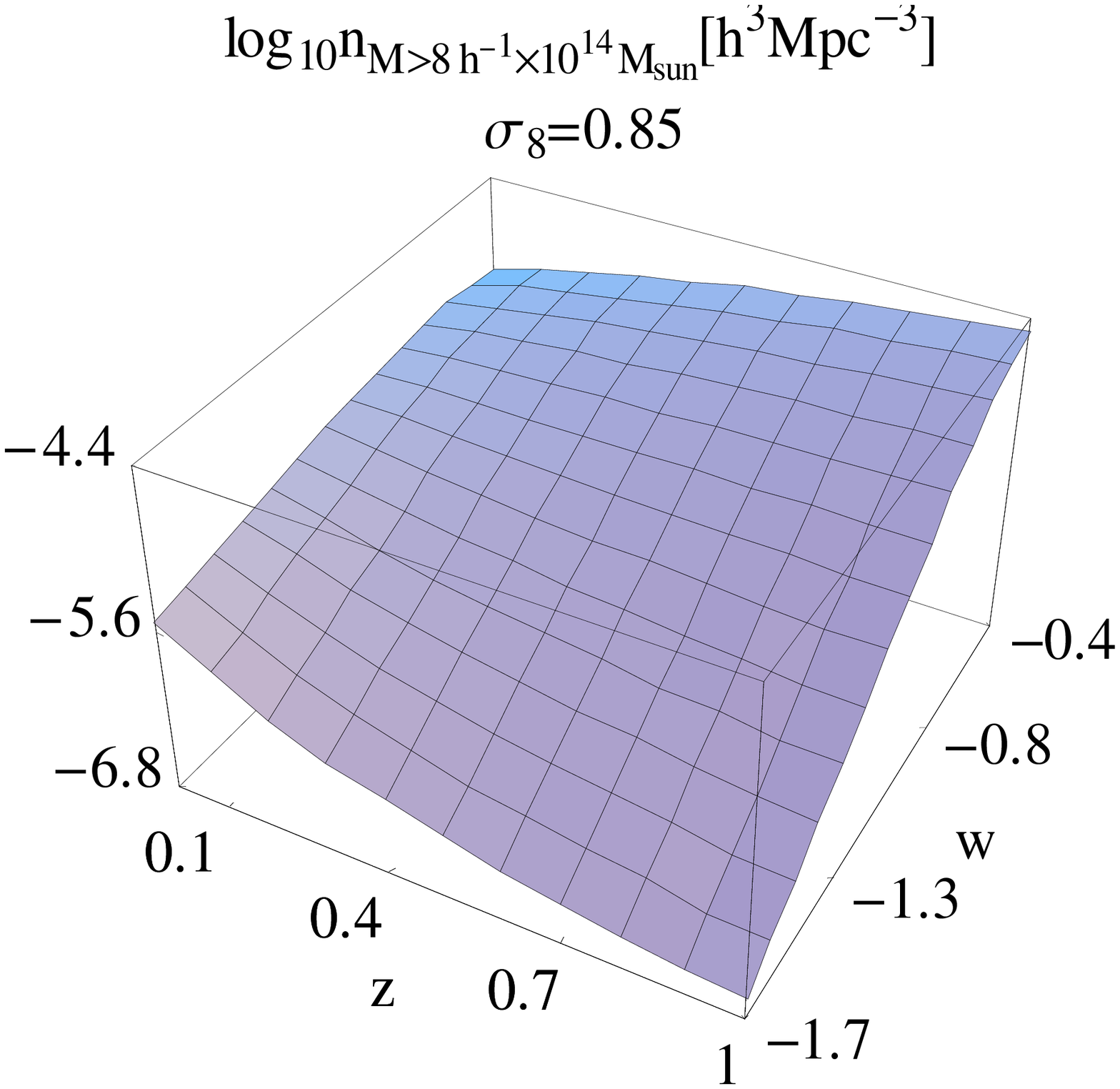}
 \caption{Effects of $w$ on the evolution of number density of
 massive galaxy clusters. The larger is $\sigma_8$, the more
 remarkable the effects of $w$ on the evolution will be.
 Both figures have $\Omega_{m0}=0.27$, $h=0.71$ and $n_s=1.0$.
 }
 \label{ew_nMfunc}
 \end{figure}

 \subsection{Numerical Results, Effects of
 $w$ on the Number Density of Galaxy Clusters and
 Its Evolutions}

 In FIG.\ref{ew_nTcore}, we display the effects
 of $w$ on the galaxy clusters
 number density v.s. temperature relation when $z=0$ for two
 values of $\frac{\mu}{\beta}$.
 From this figure we can see that the value of $w$ affects the number
 density of galaxies exponentially. Just as we explained
 in section \ref{MetricAnsaltz}, since the author of
 \cite{WangSteinhardt1} and \cite{NWeinberg} did not
 include the Q-current in their energy moment tensor
 at the right hand side of Einstein equations, which should appear
 under their assumptions, the conclusions there about the effects
 of $w$ on the number density v.s. temperature relation is
 worth further explorations. Although we do not directly correct their analysis
 by including such a current, we only assume that Quintessence
 will cluster on the scale of galaxy clusters so that such a
 current do not exist at all,
 our results here may from the contrary
 tell us the importance of such a current on the scale of galaxy clusters.

 Comparing the number density v.s. temperature relations between
 the two values of $\frac{\mu}{\beta}$ case, we can see
 that this parameter also affects the relation strongly.
 This parameter $\frac{\mu}{\beta}$ has
 physical meaning of efficiency of energy transformation
 from thermal-dynamic to x-ray.
 So the lower is this efficiency, the less possibly a galaxy
 cluster will be observed if its mass is not big enough. While the
 larger is the clusters, the less the number density of them will be.
 In \cite{ECF} and \cite{Pen},
 the authors fit the number density v.s. temperature relations to
 numerical simulations to normalize this parameter and get
 that $\frac{\mu}{\beta}\approx 1.3\sim1.4$.
 By the notations of these works, the symbol $f_{\beta}$ is used
 to denote the same relevant quantity as our $\frac{\mu}{\beta}$,
 our $f_u=1$, see the notations under eq(4) of \cite{WangSteinhardt1}.
 Considering the time dependence of mass-temperature relation
 explained by \cite{WangSteinhardt1}, the value of this parameter
 is about $0.944$ according to \cite{WangSteinhardt1}.
 Since we assume that
 Quintessence clusters synchronously with ordinary matters,
 while the Q-component could not make contributions
 to the x-ray emission processes. We
 should expect an even lower value of $\frac{\mu}{\beta}$ than $0.944$.
 In the numerical analysis in the next
 subsection, we will set $\frac{\mu}{\beta}=0.75$, but will
 point out qualitatively the way a larger or smaller value of
 $\frac{\mu}{\beta}$ affects our conclusions.

 In FIG.\ref{ew_nMfunc} we displayed the effects
 of $w$ on the evolution of number density of
 massive galaxy clusters for one value of $M_0$.
 Note that
 \beq{}
 n_{M>M_0}=\int_{M_0}^\infty n(M^\prime,z)dM^\prime
 \label{AccMfunction}
 \eeq
 From the figure we can also see that the clustering of
 Quintessence affects the evolution of the number density of
 massive galaxy clusters exponentially, and that the
 larger is $\sigma_8$, the more remarkable the effects will be.

 In this figure,
 what should interests us mostly may be, if $w$ is too greater than $-1$
 , the number density of galaxy clusters would increase as we
 look back to the past, while if $w$ is too less than $-1$,
 the contrary is the trend. And, the larger is $\sigma_8$, the
 strong this trend will be.
 Physically, this trend can be easily to understand. Because,
 the more $w\rightarrow 0$, the more weakly
 Quintessence anti-clusters, so the more usually
 emerging of galaxy clusters could take place which will lead to the
 decreasing of number of galaxy clusters. On the contrary,
 if $w<-1$, then the more $w\rightarrow -\infty$, the more strongly
 Quintessence anti-clusters, so the more often
 splitting of massive galaxy clusters will take place which
 will lead to the increasing of the number of galaxy clusters.
 According to the current observations \cite{BahcallBode03}:
 as we look back to past, we see smaller and smaller
 number density of galaxy clusters. This may implies that the
 equation of state coefficients of dark energies should not be
 greater than $-1$ too much.

 Note that, we do not directly consider the emerging and splitting
 of galaxy clusters. However, our explanations about the
 two contrary evolution trends of the number density of massive
 clusters above is still valid. Because, there are such
 configurations where initially "two" clusters are closely located, if
 $w<-1$, then as time goes on, this two clusters go far and far
 and finally really become two clusters. However, if $-1<w$, then
 as time goes on, these "two" clusters will not separate until
 today and have to be counted as one. This difference of
 "2"$\rightarrow$2 and "2"$\rightarrow$1 processes can be sensed by
 Press-Scheter theory itself through the effects of $w$ on the
 key parameter $\delta_c$ displayed in FIG.\ref{dc} with no use of
 direct consideration of galaxy clusters emerging and splitting
 processes \cite{LaceyCole, LaceyCole94, Sasaki}.

 \subsection{Constraints From Observational Results}

 Although the clustering of dark energy on the scale of
 galaxy clusters is only a logically possible
 assumption. Since this assumption predicts that
 the number density of galaxy clusters and its evolutions are affected exponentially
 by the equation of state coefficient
 $w$ of dark energy. If theoretical predictions are to be
 coincide with observations, we should have constraints on $w$,
 $\Omega_{m0}$ and $\sigma_8$ which may be rather different from
 that comes from other observation such as CMB \cite{WMAP} or
 large scale matter power spectrum \cite{SDSS}. Does these
 observation leave spaces for our assumption?

 Motivated by this fact, we try our best to fit the observation
 results of \cite{HA91} (with a factor of 2
 corrected by \cite{Henry97}) and \cite{BahcallBode03} into
 theoretical formulaes to find constrains on $w$, $\Omega_{m0}$
 and $\sigma_8$. Our
 fitting method is minimizing $\chi^2$, for details
 please refer to \cite{PressNR}. However, constrained by our
 computation powers, we do not vary all the parameters involved in
 this problem and look for the best composition.
 Instead, we choose to fit the measurements into
 theoretical formulaes in three cases. In the first case, we
 focus our attentions on $\sigma_8$ and $\Omega_{m0}$ but fix
 the other parameters. In the second case we
 focus our attention on $\sigma_8$ and $w$,
 while in the third case we focus our
 attentions on $\Omega_{m0}$ and $w$. We also tried best to fit
 the four parameters $\frac{\mu}{\beta}$, $w$, $\Omega_{m0}$ and $\sigma_8$
 simultaneously, but the convergence of the fitting
 result is so poor that we can not get any meaningful
 conclusions so we will not provide numerical results in this
 case.

 In the data source we used in this paper, \cite{HA91}
 is the observed number density v.s. temperature
 relation at $z\approx0.05$. The data from this source
 only gives errors in the $y$(number density v.s. temperature function values) axis.
 We will fit it with our theoretical
 formula eq(\ref{diff-T-Function}). \cite{BahcallBode03} is
 the observed number density v.s. red-shift relation of
 galaxy clusters whose comoving
 $1.5h^{-1}\textrm{Mpc}$ radius inside mass is
 greater than $8\times10^{14}h^{-1}M_{\odot}$. The data
 from this source have large errors in both the $x$(red-shift)
 and $y$(number density) axis. We will neglect its errors on the
 $x$ axis although it is very large. We also neglecte the
 fact that the number density provided by this source is only
 that of the clusters whose comoving $1.5h^{-1}\textrm{Mpc}$ radius
 inside mass is greater than $8\times10^{14}h^{-1}M_{\odot}$ and
 think that is the number density of clusters whose total mass is
 greater than $8\times10^{14}h^{-1}M_{\odot}$. This approximation
 will make our fitted $\sigma_8$ greater than the real value.
 For other observational results in this area, please refer to
 \cite{BFC97} and the references therein.

 For reasons explained in the above two paragraphs, we should not
 look at the numerical results of this subsection too seriously.
 What should be emphasized is the consistency between our assumption
 that dark energy will cluster synchronously with ordinary matters
 and the observational results of WMAP and SDSS.

\begin{figure}
 \includegraphics[scale=0.5]{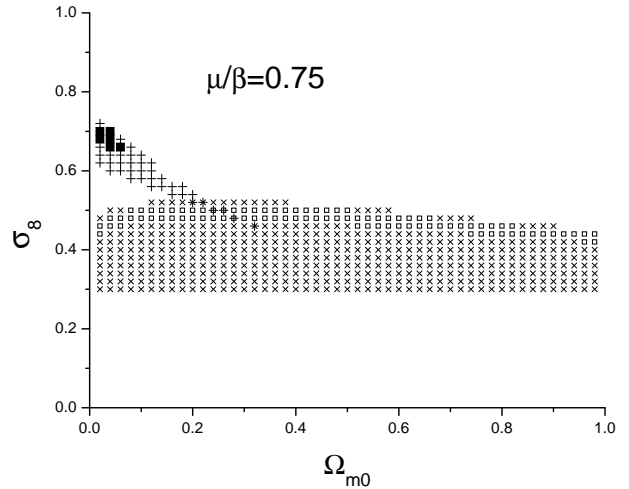}
 \caption{$\Lambda$CDM cosmology best fitting of galaxy clusters'
 observation. The black boxed region denote the $68.3\%$ confidence region
 of best fitting the number density v.s. temperature relation reported
 by \cite{HA91}, the $+$ed region denote the $99\%$ confidence
 region of the same fitting. The blank boxed denote the $68.3\%$
 confidence region of best fitting the number density v.s.
 red-shift relation reported by \cite{BahcallBode03}, while $\times$ed
 region, the $99\%$ confidence. If $\frac{\mu}{\beta}$ takes values
 greater than $0.75$, the black boxed and $+$ed region moves down
 left.
 }
 \label{Lfit}
 \end{figure}

\begin{figure}
 \includegraphics[scale=0.5]{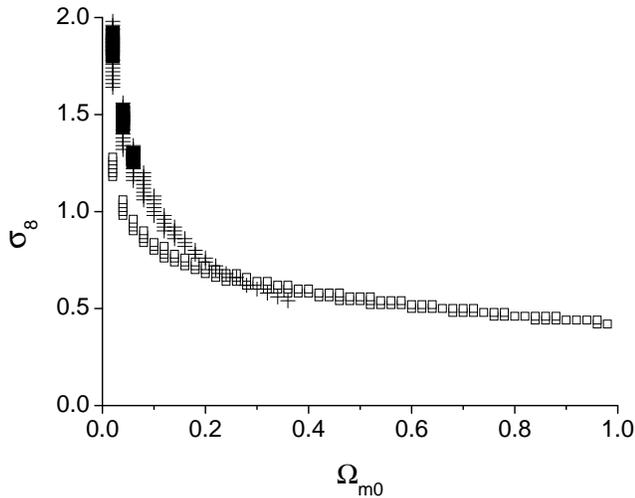}
 \caption{Still $\Lambda$CDM cosmology best fitting of galaxy clusters'
 observation, but ignored the normalization denominator $D_1(1)$ of the
 growth factor appearing in eq(\ref{sigmaRz}). The meaning of different
 symbols appearing in the figure is the same as that of FIG.\ref{Lfit}.
 }
 \label{ECFLfit}
 \end{figure}

 We display our confidence level analysis of the
 best fitting in the first case in
 FIG.\ref{Lfit} where except $\sigma_8$ and $\Omega_{m0}$, the
 other parameters are fixed as $w=-1$, $h=0.71$, $n_s=1.0$ and
 $\frac{\mu}{\beta}=0.75$. In this figure there are two points
 which is worth noticing or could be criticized by peoples.
 The first is, even use only the observed number density v.s.
 temperature relation of \cite{HA91}, we determined the cosmological
 parameters $\sigma_8$ and $\Omega_{m0}$ to some degree. This
 forms a strong contrast with the conclusion of \cite{ECF} and
 \cite{WangSteinhardt1}, which says that only using this
 observational data we cannot determine any one of this two
 parameters but only a special composition of them. We find if we
 neglect the normalization factor of the growth factor $D_1(1)$
 in eq(\ref{sigmaRz}), we
 will qualitatively reproduce the results of \cite{ECF} and
 \cite{WangSteinhardt1}, please see our FIG.\ref{ECFLfit}. We also
 find that to get the results of \cite{ECF}, Figure 1, we need to
 artificially set the initial value of either of the two
 parameters $\delta_0$ or $\kappa$, eq(A16) and (A21) of \cite{ECF}.
 The second important
 point which could be criticized by peoples is, our results indicate that
 $\sigma_8\approx0.5$, which is exceptional low comparing with the
 results of WMAP\cite{WMAP} and SDSS\cite{SDSS}. We think
 this is because when WMAP or SDSS make their best fittings, they
 did not consider the perturbation of dark energies. If the
 perturbation of dark energies are considered, we expect WMAP and
 SDSS would also predict a lower $\sigma_8$ than their
 current reported value. We note that the current version (V4.5.1) of
 the program CMBFast have been able to calculate the Cosmological
 Microwave Background Anisotropy (CMBA) spectrums when dark energies are perturbed
 , but WMAP and SDSS did not use this version of CMBFast to fit their
 results.

 As is well known, an exceptional low $\sigma_8$ implies a
 strongly
 biased universe, while $\sigma_8=1$ implies a non biased
 universe, which has been observed by the
 galaxy astronomers. So our result indicate that, on the scale of galaxy
 clusters, the mass distribution is rather different from that on
 the scale of galaxies. This is just what we should expect,
 because we assume that on the scale of galaxy clusters,
 Quintessence clusters like ordinary matters, while on the
 scale of galaxies, the clustering effects of Quintessence are usually
 neglected by peoples. Physically this is because on the scale of
 galaxy clusters, Quintessence contribute a rather large part to
 the total energy of our studying objects, while on the scale of
 galaxies, the contributions of Quintessence to the total energy
 of the studying objects is really negligible.
 It should be noted that our conclusion
 $\sigma_8\approx0.5$ is obtained by fitting the observational
 result of the number density v.s. red-shift relation into
 theoretical formulaes and has no dependence on our
 choice of $\frac{\mu}{\beta}=0.5$.

 \begin{figure}
 \includegraphics[scale=0.5]{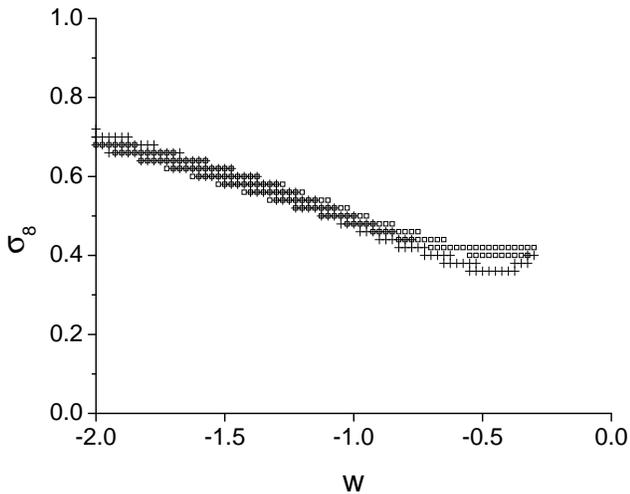}
 \caption{QCDM cosmology best fitting of the galaxy
 clusters' observation, except $w$ and $\sigma_8$, all the other
 parameters are fixed as $\Omega_{m0}=0.27$, $\frac{\mu}{\beta}=0.75$,
 $h=0.71$ and $n_s=1.0$. The $+$ed region denote the $68.3\%$
 confidence region of best fitting the number density v.s.
 temperature relation reported by \cite{HA91}, the
 blank boxed region denote the $68.3\%$ confidence region of best fitting
 the number density v.s.
 red-shift relation reported by \cite{BahcallBode03}.
 }
 \label{Qwsfit}
 \end{figure}

 \begin{figure}
 \includegraphics[scale=0.5]{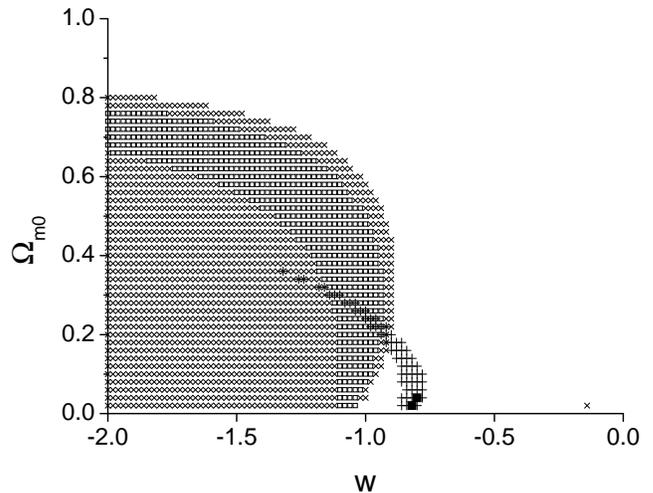}
 \caption{QCDM cosmology best fitting of the galaxy
 clusters' observation, except $w$ and $\Omega_{m0}$, all the other
 parameters are fixed as $\sigma_8=0.5$, $\frac{\mu}{\beta}=0.75$,
 $h=0.71$ and $n_s=1.0$. The black boxed region
 denote the $68.3\%$ confidence region
 of best fitting the number density v.s. temperature relation reported
 by \cite{HA91}, the $+$ed region denote the $99\%$ confidence
 region of the same fitting. The blank boxed region denote the $68.3\%$
 confidence region of best fitting the number density v.s.
 red-shift relation reported by \cite{BahcallBode03},
 while the $\times$ed region,
 the $99\%$ confidence. If $\frac{\mu}{\beta}$ takes values
 greater than $0.75$, the black boxed and $+$ed region moves down
 left.
 }
 \label{Qwofit}
 \end{figure}

 In FIG.\ref{Qwsfit}, we display the confidence level analysis
 of our best fitting in the second case where except $\sigma_8$
 and $w$ all the other parameters are fixed as $\Omega_{m0}=0.27$,
 $h=0.71$, $n_s=1.0$ and $\frac{\mu}{\beta}=0.75$. About this
 figure we would like give two comments. The first is, our
 results give an exceptional low value of $\sigma_8$, the
 physical reason of which we have explained in the previous paragraph. The
 second point is, although $w$ affects the number density v.s.
 temperature functions
 and the number density v.s. red-shift relations of massive galaxy clusters
 exponentially, see FIG.\ref{ew_nTcore} and \ref{ew_nMfunc},
 requiring theoretical predictions to coincide with observations
 does not give constraining on $w$.
 This is because, the effects of increasing $w$ and
 decreasing $\sigma_8$ on the number
 density of galaxy clusters counted each other so
 this two parameters degenerate.

 In FIG.\ref{Qwofit}, we display our confidence level analysis
 of the best fitting in the third case where except $w$
 and $\Omega_{m0}$, all the other parameters are fixed
 as $\sigma_8=0.5$,
 $h=0.71$, $n_s=1.0$ and $\frac{\mu}{\beta}=0.75$.
 On the basis of these priors, we get $w=-1.08\pm0.09$, $\Omega_{m0}=0.27\pm0.09$
 to $99\%$ confidence level. For
 reasons explained in the first two paragraph of this subsection,
 we should not look at this numerical results too
 seriously.

 \section{Conclusions}\label{ConSection}

 We studied the top-hat spherical collapse model of galaxy clusters
 formation in the flat QCDM or Phantom-CDM cosmologies under the
 assumption that Quintessence or Phantom clusters or
 anti-clusters like ordinary matters. We found that under this
 assumption, the key parameters of the model exhibit rather
 non-trivial and remarkable dependence on the equation of state
 coefficients $w$ of Quintessence or Phantoms. We then applied the
 results in Press-Scheter theory and calculated the number density
 v.s. temperature function and the evolution
 of the number density of massive galaxy clusters and found that these two
 Quantities are both affected by $w$ exponentially.

 For the number density v.s. temperature function
 of galaxy clusters, we found that the
 nearer $w\rightarrow0$, the larger the function value
 will be, the more $w\rightarrow-\infty$, the smaller the function
 value.
 While for the evolution of the number density
 of massive galaxy clusters, we found that if $w$ is too less than
 $-1$, the number density decreases as we look back to the past. On the
 contrary, if $w$ is too greater than $-1$, the number density of
 massive galaxy clusters increases as we look back.
 Using this fact, we studied the possibility of
 determining $w$ by the observational results of galaxy clusters.
 Constrained by our computation powers and the big errors of the
 observational data, we cannot determine all the
 parameters involved in this problem independently. However, on
 the basis some priors, we get that to $99\%$ confidence level,
 $w=-1.08\pm0.09$. On the other hand, if we fix $w=-1$ as priors, we find
 that $\sigma_8\approx0.5$ which is exceptional low comparing with
 the reported value of WMAP and SDSS. This is because WMAP and
 SDSS both did not consider the perturbation of dark energies.

 Since from the beginning, we
 assume that dark energy will cluster synchronously with
 ordinary matters, we have no dark energy current flowing outside
 over-dense region
 galaxy clusters. Although this is too strong an assumption,
 this assumption simplifies our discussions and makes it a
 self-consistent one. As a comparison, we point out that the
 discussions in the previous literatures which assume that
 Quintessence does not cluster on the scale of galaxy clusters but
 neglect the Q-current flowing outside the over-dense region is not
 self-consistent and the conclusions of these literatures are worth
 further explorations. Our studies do not correct this problem
 directly, but may from the contrary indicate
 the importance of such a current's effects
 on the number density of galaxy clusters and
 its evolutions.

 In \cite{SphereII}, we will dispose the assumption that dark
 energy cluster synchronously with ordinary matters and will find
 that, in that case, the equation of state coefficients of
 dark energy affects the number-density of galaxy cluster even
 more strongly!

 As a discussion, we would like to point out that the actual case
 should be, dark energy should cluster on the scale of galaxy
 clusters but not do so synchronously with ordinary
 matters. So that is a case lies between the synchronous clustering
 with ordinary matter and not clustering on the scale of galaxy
 clusters at all. What ever the actual way is, the equation of
 state coefficients should affect the number density of galaxy
 clusters remarkably. So measuring the number density of galaxy
 clusters and its evolutions is a potential effective way to
 determine the equation of state coefficients.

 \section{Acknowledgements}

 Part of the numerical computations are performed on the parallel
 computers of the Inter-discipline Center of Theoretical Studies
 of ITP, CAS, Beijing, China.

\end{document}